\documentclass{llncsheads}
\usepackage{makeidx}  
\begin{document}
\title{Complex Dynamics\\ of Autonomous Communication Networks\\
and the Intelligent Communication Paradigm}
\titlerunning{A.P. Kirilyuk: Complex Dynamics of Intelligent Communication Networks}  
\author{Andrei P. Kirilyuk}
\authorrunning{WAC 2004, http://www.autonomic-communication.org/wac/program.html}   
%
\tocauthor{Andrei Kirilyuk (Institute of Metal Physics)}
\institute{Institute of Metal Physics of the National Academy of
Sciences of Ukraine\\ 36 Vernadsky Bd, Kiev-142, Ukraine 03142\\
\email{kiril@metfiz.freenet.kiev.ua}}

\maketitle              

\begin{abstract}
Dynamics of arbitrary communication system is analysed as
unreduced interaction process. The applied generalised,
universally nonperturbative method of effective potential reveals
the phenomenon of dynamic multivaluedness of competing system
configurations forced to permanently replace each other in a
causally random order, which leads to universally defined
dynamical chaos, complexity, fractality, self-organisation, and
adaptability. We demonstrate the origin of huge, exponentially
high efficiency of the unreduced, complex network dynamics and
specify the universal symmetry of complexity as the fundamental
guiding principle for creation and control of such qualitatively
new kind of networks and devices.
\end{abstract}
\section{Introduction}
Any communication system can be considered as a particular case of
general dynamical system formed by many interacting units. If the
system components are permitted to freely interact without
\emph{strict} external control, then such unreduced interaction
process leads inevitably to complex-dynamical, essentially
nonlinear and chaotic structure emergence, or generalised
(dynamically multivalued) self-organisation
\cite{kir:1,kir:2,kir:3}, extending the conventional, basically
regular self-organisation concept. The usual technology and
communication practice and paradigm rely, however, on very strong
human control and totally regular, predictable dynamics of
controlled systems and environment, where unpredictable events can
only take the form of undesirable failures or noise.

Growing volumes and complication of communication system links and
functions lead inevitably to increasing probability of undesirable
deviations from the pre-programmed regular behaviour, largely
compromising its supposed advantages. On the other hand, such
increasingly useful properties as intrinsic system creativity and
autonomous adaptability to changing environment and individual
user demands should certainly involve another, much less regular
and more diverse kind of behaviour. In this paper we analyse these
issues in a rigorous way by presenting the unreduced,
nonperturbative analysis of an arbitrary system of interacting
entities and show that such \emph{unreduced interaction process}
possesses the natural, dynamically derived properties of
chaoticity, creativity (autonomous structure formation ability),
adaptability, and exponentially high efficiency, which can be
consistently unified into the totally universal concept of
\emph{dynamic complexity} \cite{kir:1}. This concept and
particular notions it unifies represent essential extension with
respect to respective results of the usual theory always using one
or another version of perturbation theory that strongly reduces
real interaction processes and leads inevitably to regular kind of
dynamics (even in its versions of chaoticity). We shall specify
these differences in our analysis and demonstrate the key role of
unreduced, interaction-driven complexity, chaoticity and
self-organisation in the superior operation properties, as it has
already been demonstrated for a large scope of applications
\cite{kir:1,kir:2,kir:3,kir:4,kir:5,kir:6,kir:7,kir:8}.

We start, in Sect. 2, with a mathematical demonstration of the
fact that the unreduced interaction process within \emph{any} real
system leads to intrinsic, genuine, and omnipresent
\emph{randomness} in the system behaviour, which can be realised
in a few characteristic regimes and leads to the
\emph{universally} defined \emph{dynamic complexity}. We outline
the change in strategy and practice of communication system
construction and use, which follows from such unreduced analysis
of system interactions. The \emph{universality} of our analysis is
of special importance here, since the results can be applied at
various \emph{naturally entangled} levels of communication system
operation. In particular, we demonstrate the complex-dynamic
origin of the huge, \emph{exponentially high efficiency growth} of
the unreduced, causally random system dynamics, with respect to
the standard, basically regular system operation (Sect. 3).
Finally, the dynamically derived, \emph{universal symmetry, or
conservation, of complexity} is introduced as the new guiding
principle and tool of complex system dynamics that should replace
usual, regular programming. The \emph{paradigm of intelligent
communication systems} is thus specified, since we show also
\cite{kir:1,kir:5} that the property of \emph{intelligence} can be
consistently described as high enough levels of the unreduced
dynamic complexity. This ``intelligent communication" is the most
complete, inevitable realisation, and in fact a synonym, of the
truly \emph{autonomous} communication dynamics and its expected
properties.
\section{Complex Dynamics of Unreduced Interaction Process}
We begin with a general expression of multi-component system
dynamics (or many-body problem), called here \emph{existence
equation}, fixing the fact of interaction between the system
components, and generalising various model equations:
\begin{equation}\label{eq:1}
\left\{ {\sum\limits_{k = 0}^N {\left[ {h_k \left( {q_k } \right)
+ \sum\limits_{l > k}^N {V_{kl} \left( {q_k ,q_l } \right)} }
\right]} } \right\}\Psi \left( Q \right) = E\Psi \left( Q \right)\
,
\end{equation}
where $h_k \left( {q_k } \right)$ is the ``generalised
Hamiltonian" of the $k$-th system component in the absence of
interaction, $q_k$ is the degree(s) of freedom of the $k$-th
component (expressing its ``physical nature"), $V_{kl} \left( {q_k
,q_l } \right)$ is the (generally arbitrary) interaction potential
between the $k$-th and $l$-th components, $\Psi \left( Q \right)$
is the system state-function, $Q \equiv \left\{ {q_0 ,q_1 ,...,q_N
} \right\}$, $E$ is the eigenvalue of the generalised Hamiltonian,
and summations are performed over all ($N$) system components. The
generalised Hamiltonian, eigenvalues, and interaction potential
represent a suitable measure of dynamic complexity defined below
and encompassing practically all ``observable" quantities (action,
energy, momentum, current, etc.) at any level of dynamics.
Therefore (\ref{eq:1}) can express the unreduced interaction
configuration at any level of communication network of arbitrary
initial structure. It can also be presented in a particular form
of time-dependent equation by replacing the generalised
Hamiltonian eigenvalue $E$ with the partial time derivative
operator (for the case of explicit interaction potential
dependence on time).

One can separate one of the degrees of freedom, e.g. $q_0 \equiv
\xi $, corresponding to a naturally selected, usually
``system-wide" entity, such as ``embedding" configuration (system
of coordinates) or common ``transmitting agent":
\begin{equation}\label{eq:2}
\left\{ {h_0 \left( \xi  \right) + \sum\limits_{k = 1}^N {\left[
{h_k \left( {q_k } \right) + V_{0k} \left( {\xi ,q_k } \right)}
\right] + } \sum\limits_{l > k}^N {V_{kl} \left( {q_k ,q_l }
\right)} } \right\}\Psi \left( {\xi ,Q} \right) = E\Psi \left(
{\xi ,Q} \right),
\end{equation}
where now $Q \equiv \left\{ {q_1 ,...,q_N } \right\}$ and $k,l \ge
1$.

We then express the problem in terms of known free-component
solutions for the ``functional", internal degrees of freedom of
system elements ($k \ge 1$):
\begin{equation}\label{eq:3}
h_k \left( {q_k } \right)\varphi _{kn_k } \left( {q_k } \right) =
\varepsilon _{n_k } \varphi _{kn_k } \left( {q_k } \right)\ ,
\end{equation}
\begin{equation}\label{eq:4}
\Psi \left( {\xi ,Q} \right) = \sum\limits_n {\psi _n \left( \xi
\right)} \varphi _{1n_1 } \left( {q_1 } \right)\varphi _{2n_2 }
\left( {q_2 } \right)...\varphi _{Nn_N } \left( {q_N } \right)
\equiv \sum\limits_n {\psi _n \left( \xi  \right)} \Phi _n \left(
Q \right),
\end{equation}
where $\left\{ {\varepsilon _{n_k } } \right\}$ are the
eigenvalues and $\left\{ {\varphi _{kn_k } \left( {q_k } \right)}
\right\}$ eigenfunctions of the $k$-th component Hamiltonian $h_k
\left( {q_k } \right)$, forming the complete set of orthonormal
functions, $n \equiv \left\{ {n_1,...,n_N } \right\}$ runs through
all possible eigenstate combinations, and $\Phi _n \left( Q
\right) \equiv \varphi _{1n_1 } \left( {q_1 } \right)\varphi
_{2n_2 } \left( {q_2 } \right)...\varphi _{Nn_N } \left( {q_N }
\right)$ by definition. The system of equations for $\left\{ {\psi
_n \left( \xi  \right)} \right\}$ is obtained then in a standard
way, using the eigen-solution orthonormality (e.g. by
multiplication by $\Phi _n^
* \left( Q \right)$ and integration over $Q$):
\begin{equation}\label{eq:5}
\begin{array}{rcl}
\left[ {h_0 \left( \xi  \right) + V_{00} \left( \xi  \right)}
\right]\psi _0 \left( \xi  \right)&+&\sum\limits_n {V_{0n} \left(
\xi  \right)} \psi _n \left( \xi  \right) = \eta \psi _0 \left(
\xi  \right)\\
\left[ {h_0 \left( \xi  \right) + V_{nn} \left( \xi \right)}
\right]\psi _n \left( \xi  \right)&+&\sum\limits_{n' \ne n}
{V_{nn'} \left( \xi  \right)} \psi _{n'} \left( \xi  \right) =
\eta _n \psi _n \left( \xi  \right) - V_{n0} \left( \xi
\right)\psi _0 \left( \xi  \right),
\end{array}
\end{equation}
where $n,n' \ne 0$ (also below), $\eta \equiv \eta _0 = E -
\varepsilon _0$, $\eta _n = E - \varepsilon _n$, $\varepsilon _n =
\sum\limits_k {\varepsilon _{n_k } }$,
\begin{equation}\label{eq:6}
V_{nn'} \left( \xi  \right) = \sum\limits_k {\left[ {V_{k0}^{nn'}
\left( \xi  \right) + \sum\limits_{l > k} {V_{kl}^{nn'} } }
\right]}\ ,
\end{equation}
\begin{equation}\label{eq:7}
V_{k0}^{nn'} \left( \xi  \right) = \int\limits_{\Omega _Q }
{dQ\Phi _n^ *  \left( Q \right)} V_{k0} \left( {q_k ,\xi }
\right)\Phi _{n'} \left( Q \right)\ ,
\end{equation}
\begin{equation}\label{eq:8}
V_{kl}^{nn'} \left( \xi \right) = \int\limits_{\Omega _Q } {dQ\Phi
_n^ *  \left( Q \right)} V_{kl} \left( {q_k ,q_l } \right)\Phi
_{n'} \left( Q \right)\ ,
\end{equation}
and we have separated the equation for $\psi _0 \left( \xi
\right)$ describing the generalised ``ground state" of the system
elements, i. e. the state with minimum complexity. The obtained
system of equations expresses the same problem as the starting
equation (\ref{eq:2}), but now in terms of ``natural", dynamic
variables, and therefore it can be obtained for various starting
models, including time-dependent and formally ``nonlinear" ones
(see below for a rigorous definition of \emph{essential}
nonlinearity).

We try now to approach the solution of the ``nonintegrable" system
of equations (\ref{eq:5}) with the help of the generalised
effective, or optical, potential method \cite{ded}, where one
expresses $\psi _n \left( \xi  \right)$ through $\psi _0 \left(
\xi \right)$ from the equations for $\psi _n \left( \xi  \right)$
using the standard Green function technique and then inserts the
result into the equation for $\psi _0 \left( \xi  \right)$,
obtaining thus the \emph{effective existence equation} that
contains \emph{explicitly} only ``integrable" degrees of freedom
($\xi$) \cite{kir:1,kir:2,kir:3,kir:4,kir:5,kir:6,kir:7,kir:8}:
\begin{equation}\label{eq:9}
h_0 \left( \xi  \right)\psi _0 \left( \xi  \right) +
V_{{\rm{eff}}} \left( {\xi ;\eta } \right)\psi _0 \left( \xi
\right) = \eta \psi _0 \left( \xi  \right)\ ,
\end{equation}
where the operator of \emph{effective potential (EP)},
$V_{{\rm{eff}}} \left( {\xi ;\eta } \right)$, is given by
\begin{equation}\label{eq:10}
V_{{\rm{eff}}} \left( {\xi ;\eta } \right) = V_{00} \left( \xi
\right) + \hat V\left( {\xi ;\eta } \right),\ \  \hat V\left( {\xi
;\eta } \right)\psi _0 \left( \xi  \right) = \int\limits_{\Omega
_\xi  } {d\xi 'V\left( {\xi ,\xi ';\eta } \right)} \psi _0 \left(
{\xi '} \right),
\end{equation}
\begin{equation}\label{eq:11}
V\left( {\xi ,\xi ';\eta } \right) = \sum\limits_{n,i}
{\frac{{V_{0n} \left( \xi  \right)\psi _{ni}^0 \left( \xi
\right)V_{n0} \left( {\xi '} \right)\psi _{ni}^{0*} \left( {\xi '}
\right)}}{{\eta  - \eta _{ni}^0  - \varepsilon _{n0} }}}\ ,\ \ \
\varepsilon _{n0}  \equiv \varepsilon _n  - \varepsilon _0\ ,
\end{equation}
and $\left\{ {\psi _{ni}^0 \left( \xi  \right)} \right\}$,
$\left\{ {\eta _{ni}^0 } \right\}$ are complete sets of
eigenfunctions and eigenvalues of a \emph{truncated} system of
equations:
\begin{equation}\label{eq:12}
\left[ {h_0 \left( \xi  \right) + V_{nn} \left( \xi  \right)}
\right]\psi _n \left( \xi  \right) + \sum\limits_{n' \ne n}
{V_{nn'} \left( \xi  \right)} \psi _{n'} \left( \xi  \right) =
\eta _n \psi _n \left( \xi  \right)\ .
\end{equation}

One should use now the eigenfunctions, $\left\{ {\psi _{0i} \left(
\xi \right)} \right\}$, and eigenvalues, $\left\{ {\eta _i }
\right\}$, of the formally ``integrable" equation (\ref{eq:9}) to
obtain other state-function components:
\begin{equation}\label{eq:13}
\psi _{ni} \left( \xi  \right) = \hat g_{ni} \left( \xi
\right)\psi _{0i} \left( \xi  \right) \equiv \int\limits_{\Omega
_\xi  } {d\xi 'g_{ni} \left( {\xi ,\xi '} \right)\psi _{0i} \left(
{\xi '} \right)}\ ,
\end{equation}
\begin{equation}\label{eq:14}
g_{ni} \left( {\xi ,\xi '} \right) = V_{n0} \left( {\xi '}
\right)\sum\limits_{i'} {\frac{{\psi _{ni'}^0 \left( \xi
\right)\psi _{ni'}^{0*} \left( {\xi '} \right)}}{{\eta _i  - \eta
_{ni'}^0  - \varepsilon _{n0} }}}\ ,
\end{equation}
and the total system state-function, $\Psi \left( {q_0 ,q_1
,...,q_N } \right) = \Psi \left( {\xi ,Q} \right)$ (see
(\ref{eq:4})):
\begin{equation}\label{eq:15}
\Psi \left( {\xi ,Q} \right) = \sum\limits_i {c_i } \left[ {\Phi
_0 \left( Q \right) + \sum\limits_n {\Phi _n } \left( Q
\right)\hat g_{ni} \left( \xi  \right)} \right]\psi _{0i} \left(
\xi  \right)\ ,
\end{equation}
where the coefficients $c_i$ should be found from the
state-function matching conditions at the boundary where
interaction effectively vanishes. The measured quantity,
generalised as structure density $\rho \left( {\xi ,Q} \right)$,
is obtained as the state-function squared modulus, $\rho \left(
{\xi ,Q} \right) = \left| {\Psi \left( {\xi ,Q} \right)} \right|^2
$ (for ``wave-like" complexity levels), or as the state-function
itself, $\rho \left( {\xi ,Q} \right) = \Psi \left( {\xi ,Q}
\right)$ (for ``particle-like" structures) \cite{kir:1}.

Since the EP expression in the effective problem formulation
(\ref{eq:9})-(\ref{eq:11}) depends essentially on the
eigen-solutions to be found, the problem remains ``nonintegrable"
and formally equivalent to the initial formulation (\ref{eq:1}),
(\ref{eq:2}), (\ref{eq:5}). However, it is the effective version
of a problem that leads to its unreduced solution and reveals the
nontrivial properties of the latter
\cite{kir:1,kir:2,kir:3,kir:4,kir:5,kir:6,kir:7,kir:8}. The most
important property of the unreduced interaction result
(\ref{eq:9})-(\ref{eq:15}) is its \emph{dynamic multivaluedness}
meaning that one has a \emph{redundant} number of different but
individually complete, and therefore \emph{mutually incompatible},
problem solutions, each of them describing an \emph{equally real}
system configuration. We call each such locally complete solution
(and real system configuration) \emph{realisation} of the system
and problem. Plurality of system realisations follows from the
unreduced EP expressions due to the nonlinear and self-consistent
dependence on the solutions to be found, reflecting the physically
real and evident plurality of possible combinations of interacting
eigen-modes
\cite{kir:1,kir:2,kir:3,kir:4,kir:5,kir:6,kir:7,kir:8}. It is
important that dynamic multivaluedness emerges only in the
unreduced problem formulation, whereas the standard theory,
including EP method applications (see e.g. \cite{ded}) and the
scholar ``science of complexity" (theory of chaos,
self-organisation, etc.), resorts invariably to one or another
version of perturbation theory, whose approximation, used to
obtain an ``exact", closed-form solution, totally ``kills"
redundant solutions by eliminating just those nonlinear dynamical
links and retains \emph{only one}, ``averaged" solution, usually
expressing only \emph{small} deviations from initial,
pre-interaction configuration. This \emph{dynamically
single-valued}, or \emph{unitary}, problem reduction forms the
basis of the whole canonical science paradigm.

Since we have many \emph{incompatible} system realisations that
tend to appear from the same, driving interaction, we obtain the
key property of \emph{causal, or dynamic, randomness} in the form
of permanently \emph{changing} realisations that replace each
other in the \emph{truly random} order. Therefore dynamic
multivaluedness, rigorously derived simply by unreduced, correct
solution of a real many-body (interaction) problem, provides the
\emph{universal dynamic origin} and \emph{meaning} of the
\emph{omnipresent, unceasing} randomness in the system behaviour,
also called \emph{(dynamical) chaos} (it is essentially different
from any its unitary version, reduced to an ``involved regularity"
or \emph{postulated} external ``noise"). This means that the
genuine, truly complete \emph{general solution} of an arbitrary
problem (describing a \emph{real} system behaviour) has the form
of \emph{dynamically probabilistic} sum of measured quantities for
particular system realisations:
\begin{equation}\label{eq:16}
\rho \left( {\xi ,Q} \right) = \sum\limits_{r = 1}^{N_\Re  } {^{^
\oplus}  \rho _r \left( {\xi ,Q} \right)}\ ,
\end{equation}
where summation is performed over all system realisations, $N_\Re$
is their number (its maximum value is equal to the number of
system components, $N_\Re = N$), and the sign $\oplus$ designates
the special, dynamically probabilistic meaning of the sum
described above. It implies that any measured quantity
(\ref{eq:16}) is \emph{intrinsically unstable} and its current
value \emph{will} unpredictably change to another one,
corresponding to another, \emph{randomly} chosen realisation. Such
kind of behaviour is readily observed in nature and actually
explains the living organism behaviour \cite{kir:1,kir:4,kir:5},
but is thoroughly avoided in the unitary theory and technological
systems (including communication networks), where it is correctly
associated with linear ``noncomputability" and technical failure
(we shall consider below this \emph{limiting} regime of real
system dynamics). Therefore the universal dynamic multivaluedness
thus revealed by the rigorous problem solution forms the
fundamental basis for the transition to ``bio-inspired" and
``intelligent" kind of operation in artificial, technological and
communication systems, where causal randomness can be transformed
from an obstacle to a qualitative advantage (Sect. 3).

The rigorously derived randomness of the generalised EP formalism
(\ref{eq:9})-(\ref{eq:16}) is accompanied by the \emph{dynamic
definition of probability}. Because the elementary realisations
are equivalent in their ``right to appear", the dynamically
obtained, \emph{a priori probability}, $\alpha _r$, of an
elementary realisation emergence is given by
\begin{equation}\label{eq:17}
\alpha _r  = \frac{1}{{N_\Re  }}\ ,\ \ \ \sum\limits_r {\alpha _r
} = 1\ .
\end{equation}
However, a real observation may fix uneven groups of elementary
realisations because of their multivalued self-organisation (see
below). Therefore the dynamic probability of observation of such
general, compound realisation is determined by the number, $N_r$,
of elementary realisations it contains:
\begin{equation}\label{eq:18}
\alpha _r \left( {N_r } \right) = \frac{{N_r }}{{N_\Re  }}\ \ \
\left( {N_r  = 1,...,N_\Re  ;\ \sum\limits_r {N_r }  = N_\Re  }
\right),\ \ \ \sum\limits_r {\alpha _r }  = 1\ .
\end{equation}
An expression for \emph{expectation value}, $\rho _{\exp } \left(
{\xi ,Q} \right)$, can easily be constructed from
(\ref{eq:16})-(\ref{eq:18}) for statistically long observation
periods:
\begin{equation}\label{eq:19}
\rho _{\exp } \left( {\xi ,Q} \right) = \sum\limits_r {\alpha _r
\rho _r \left( {\xi ,Q} \right)}\ .
\end{equation}
It is important, however, that our dynamically derived randomness
and probability need not rely on such ``statistical", empirically
based result, so that the basic expressions
(\ref{eq:16})-(\ref{eq:18}) remain valid even for a \emph{single}
event of realisation emergence and \emph{before} any event happens
at all.

The realisation probability distribution can be obtained in
another way, involving \emph{generalised wavefunction} and
\emph{Born's probability rule}
\cite{kir:1,kir:3,kir:5,kir:8,kir:9}. The wavefunction describes
the system state during its transition between ``regular",
``concentrated" realisations and constitutes a particular,
``intermediate" realisation with spatially extended and ``loose"
(chaotically changing) structure, where the system components
transiently disentangle before forming the next ``regular"
realisation. The intermediate, or ``main", realisation is
explicitly obtained in the unreduced EP formalism
\cite{kir:1,kir:3,kir:5,kir:8,kir:9} and provides, in particular,
the causal, totally realistic version of the quantum-mechanical
wavefunction at the lowest, ``quantum" levels of complexity. The
``Born probability rule", now also causally derived and extended
to any level of world dynamics, states that the realisation
probability distribution is determined by the wavefunction values
(their squared modulus for the ``wave-like" complexity levels) for
the respective system configurations. The generalised wavefunction
(or distribution function) satisfies the universal Schr\"odinger
equation (Sect. 3), rigorously derived from the dynamic
quantization of complex dynamics
\cite{kir:1,kir:3,kir:5,kir:8,kir:9}, while Born's probability
rule follows from the \emph{dynamic} ``boundary conditions"
mentioned in connection to the state-function expression
(\ref{eq:15}) and actually satisfied just during each system
transition between a ``regular" realisation and the extended
wavefunction state. Note also that it is this ``averaged",
weak-interaction state of the wavefunction, or ``main"
realisation, that actually remains in the dynamically
single-valued, one-realisation ``model" and ``exact-solution"
paradigm of the unitary theory, which explains both its partial
success and fundamental limitations.

Closely related to the dynamic multivaluedness is the property of
\emph{dynamic entanglement} between the interacting components,
described in (\ref{eq:15}) by the dynamically weighted products of
state-function components depending on various degrees of freedom
($\xi, Q$). It provides a rigorous expression of the tangible
\emph{quality} of the emerging system structure and is absent in
unitary models. The obtained \emph{dynamically multivalued
entanglement} describes a ``living" structure, permanently
changing and probabilistically \emph{adapting} its configuration,
which provides a well-specified basis for ``bio-inspired"
technological solutions. The properties of dynamically multivalued
entanglement and adaptability are further amplified due to the
extended \emph{probabilistic fractality} of the unreduced general
solution \cite{kir:1,kir:4,kir:5}, obtained by application of the
same EP method to solution of the truncated system of equations
(\ref{eq:12}) used in the first-level EP expression (\ref{eq:11}).

We can now consistently and universally define the unreduced
\emph{dynamic complexity}, $C$, of any real system (or interaction
process) as arbitrary growing function of the total number of
\emph{explicitly obtained} system realisations, $C = C\left(
{N_\Re  } \right),\ \ {{dC} \mathord{\left/ {\vphantom {{dC}
{dN_\Re   > 0}}} \right. \kern-\nulldelimiterspace} {dN_\Re   >
0}}$, or the rate of their change, equal to zero for the
unrealistic case of only one system realisation, $C\left( {\rm{1}}
\right){\rm{  = 0}}$. Suitable examples are provided by $C\left(
{N_\Re  } \right) = C_0 \ln N_\Re$, generalised energy/mass
(proportional to the temporal rate of realisation change), and
momentum (proportional to the spatial rate of realisation
emergence) \cite{kir:1,kir:5,kir:8,kir:9}. It becomes clear now
that the whole \emph{dynamically single-valued} paradigm and
results of the canonical theory (including its versions of
``complexity" and \emph{imitations} of ``multi-stability" in
\emph{abstract}, mathematical ``spaces") correspond to exactly
\emph{zero} value of the unreduced dynamic complexity, which is
equivalent to the effectively zero-dimensional, point-like
projection of reality in the ``exact-solution" perspective.

Correspondingly, \emph{any} dynamically single-valued ``model" is
strictly regular and \emph{cannot} possess any true, intrinsic
randomness (chaoticity), which should instead be introduced
artificially (and inconsistently), e.g. as a \emph{regular}
``amplification" of a ``random" (by convention) \emph{external}
``noise" or ``measurement error". By contrast, our unreduced
dynamic complexity is practically synonymous to the equally
universally defined and genuine \emph{chaoticity} (see above),
since multiple system realisations, appearing and disappearing
only in the \emph{real} space (and \emph{forming} thus its
tangible, changing structure \cite{kir:1,kir:3,kir:5,kir:8}), are
redundant (mutually incompatible), which is the origin of
\emph{both} complexity and chaoticity. The genuine dynamical chaos
thus obtained has its complicated internal structure (contrary to
the ill-defined unitary ``stochasticity") and always contains
\emph{partial regularity}, which is dynamically, inseparably
entangled with truly random elements.

The universal dynamic complexity, chaoticity, and related
properties involve the \emph{essential, or dynamic, nonlinearity}
of the unreduced problem solution and corresponding system
behaviour. It is provided by the naturally formed dynamical links
of the developing interaction process, as they are expressed in
the (eventually fractal) EP dependence on the problem solutions to
be found (see (\ref{eq:9})-(\ref{eq:11})). It is the
\emph{dynamically emerging} nonlinearity, since it appears even
for a formally ``linear" initial problem expression
(\ref{eq:1})-(\ref{eq:2}), (\ref{eq:5}), whereas the usual,
mechanistic ``nonlinearity" is but a perturbative approximation to
the essential nonlinearity of the unreduced EP expressions. The
essential nonlinearity leads to the irreducible \emph{dynamic
instability} of any system state (realisation), since both are
determined by the same dynamic feedback mechanism.

Universality of our description leads, in particular, to the
unified understanding of the whole diversity of existing dynamical
regimes and types of system behaviour \cite{kir:1,kir:2,kir:5}.
One standard, limiting case of complex (multivalued) dynamics,
called \emph{uniform, or global, chaos}, is characterised by
sufficiently different realisations with a homogeneous
distribution of probabilities (i.e. $N_r \approx 1$) and $\alpha
_r  \approx {1 \mathord{\left/ {\vphantom {1 {N_\Re  }}} \right.
\kern-\nulldelimiterspace} {N_\Re  }}$ for all $r$ in
(\ref{eq:18})) and is obtained when the major parameters of
interacting entities (suitably represented by frequencies) are
similar to each other (which leads to a ``strong conflict of
interests" and resulting ``deep disorder"). The complementary
limiting regime of \emph{multivalued self-organisation, or
self-organised criticality (SOC)} emerges for sufficiently
different parameters of interacting components, so that a small
number of relatively rigid, low-frequency components ``enslave" a
hierarchy of high-frequency and rapidly changing, but
configurationally similar, realisations (i.e. $N_r \sim N_\Re$ and
realisation probability distribution is highly inhomogeneous). The
difference of this extended, multivalued self-organisation (and
SOC) from the usual, unitary version is essential: despite the
rigid \emph{external} shape of the system configuration in this
regime, it contains the intense ``internal life" and \emph{chaos}
of permanently changing ``enslaved" realisations (which are
\emph{not} superposable unitary ``modes"). Another important
advance with respect to the unitary ``science of complexity" is
that the unreduced, multivalued self-organisation unifies the
extended versions of a whole series of separated unitary
``models", including SOC, various versions of ``synchronisation",
``control of chaos", ``attractors", and ``mode locking". All the
intermediate dynamic regimes between those two limiting cases of
uniform chaos and multivalued SOC (as well as their multi-level,
fractal combinations) are obtained for intermediate values of
interaction parameters. The point of transition to the strong
chaos is expressed by the \emph{universal criterion of global
chaos onset}:
\begin{equation}\label{eq:20}
\kappa  \equiv \frac{{\Delta \eta _i }}{{\Delta \eta _n }} =
\frac{{\omega _\xi  }}{{\omega _q }} \cong 1\ ,
\end{equation}
where $\kappa$ is the introduced \emph{chaoticity} parameter,
$\Delta \eta _i$, $\omega _\xi$ and $\Delta \eta _n \sim \Delta
\varepsilon$, $\omega _q$ are energy-level separations and
frequencies for the inter-component and intra-component motions,
respectively. At $\kappa \ll 1$ one has the externally regular
multivalued SOC regime, which degenerates into global chaos as
$\kappa$ grows from 0 to 1, and the maximum irregularity at
$\kappa \approx 1$ is again transformed into a multivalued SOC
kind of structure at $\kappa \gg 1$ (but with a ``reversed" system
configuration).

One can compare this transparent and universal picture with the
existing diversity of separated and incomplete unitary criteria of
chaos and regularity. Only the former provide a real possibility
of understanding and control of communication tools of arbitrary
complexity, where more regular regimes can serve for desirable
direction of communication dynamics, while less regular ones will
play the role of efficient search and adaptation means. This
combination forms the basis of any ``biological" and
``intelligent" kind of behaviour \cite{kir:1,kir:4,kir:5} and
therefore can constitute the essence of the \emph{intelligent
communication paradigm} supposed to extend the now realised
(quasi-) regular kind of communication, which corresponds to the
uttermost limit of SOC ($\kappa \to 0$). While the latter
\emph{inevitably} becomes inefficient with growing network
sophistication (where the chaos-bringing resonances of
(\ref{eq:20}) \emph{cannot} be avoided any more), it definitely
lacks the ``intelligent power" of unreduced complex dynamics to
generate meaning and adaptable structure development.
\section{Huge efficiency of complex communication dynamics
and the guiding role of the symmetry of complexity}
The \emph{dynamically probabilistic fractality} of the system
structure emerges naturally by the unreduced interaction
development itself \cite{kir:1,kir:4,kir:5}. It is obtained
mathematically by application of the same EP method
(\ref{eq:9})-(\ref{eq:14}) to solution of the truncated system of
equations (\ref{eq:12}), then to solution of the next truncated
system, etc., which gives the irregular and
\emph{probabilistically moving} hierarchy of realisations,
containing the intermittent mixture of global chaos and
multivalued SOC, which constitute together a sort of
\emph{confined chaos}. The total realisation number $N_\Re$, and
thus the power, of this autonomously branching interaction process
with a \emph{dynamically parallel} structure grows
\emph{exponentially} within any time period. It can be estimated
in the following way \cite{kir:5}.

If our system of inter-connected elements contains
$N_{{\rm{unit}}}$ ``processing units", or ``junctions", and if
each of them has $n_{{\rm{conn}}}$ real or ``virtual" (possible)
links, then the total number of interaction links is $N =
n_{{\rm{conn}}} N_{{\rm{unit}}}$. In most important cases $N$ is a
huge number: for both human brain and genome interactions $N$ is
greater than $10^{12}$, and being much more variable for
communication systems, it will typically scale in similar
``astronomical" ranges. The key property of \emph{unreduced,
complex} interaction dynamics, distinguishing it from any unitary
version, is that the maximum number $N_\Re$ of realisations
actually taken by the system (also per time unit) and determining
its real ``power" $P_{{\rm{real}}}$ (of search, memory, cognition,
etc.) is given by the number of \emph{all possible combinations of
links}, i.e.
\begin{equation}\label{eq:21}
P_{{\rm{real}}}  \propto N_\Re = N! \to \sqrt {2{\rm{\pi }}N}
\left( {\frac{N}{e}} \right)^N  \sim N^N  \gg  \gg N\ .
\end{equation}
Any unitary, sequential model of the same system (including its
\emph{mechanistically} ``parallel" and ``complex" modes) would
give $P_{{\rm{reg}}} \sim N^\beta$, with $\beta \sim 1$, so that
\begin{equation}\label{eq:22}
P_{{\rm{real}}}  \sim \left( {P_{{\rm{reg}}} } \right)^N  \gg  \gg
P_{{\rm{reg}}}  \sim N^\beta\ .
\end{equation}
Thus, for $N \sim 10^{12}$ we have $P_{{\rm{real}}} \gg
10^{10^{13} } \gg 10^{10^{12} } \sim 10^N  \to \infty $, which is
indeed a ``practical infinity", also with respect to the unitary
power of $N^\beta \sim 10^{12}$.

These estimates demonstrate the true power of complex
(multivalued) communication dynamics that remains suppressed
within the unitary, quasi-regular operation mode dominating now in
man-made technologies. The huge power values for complex-dynamical
interaction correlate with the new \emph{quality} emergence, such
as \emph{intelligence} and \emph{consciousness} (at higher levels
of complexity) \cite{kir:5}, which has a direct relation to our
\emph{intelligent} communication paradigm, meaning that such
properties as \emph{sensible}, context-related information
processing, personalised \emph{understanding} and autonomous
\emph{creativity} (useful self-development), desired for the new
generation networks, are inevitable \emph{qualitative}
manifestations of the above ``infinite" power.

Everything comes at a price, however, and a price to pay for the
above qualitative advantages is rigorously specified now as
irreducible \emph{dynamic randomness}, and thus unpredictability
of operation details in complex information-processing systems. We
only rigorously confirm here an evident conclusion that
\emph{autonomous} adaptability and genuine \emph{creativity}
exclude any detailed, regular, predictable pre-programming in
principle. But what then can serve as a guiding principle and
practical strategy of construction of those qualitatively new
types of communications networks and their ``intelligent"
elements? We show in our further analysis of complex-dynamic
interaction process that those guiding rules and strategy are
determined by a general law of complex (multivalued) dynamics, in
the form of \emph{universal symmetry, or conservation, of
complexity} \cite{kir:1,kir:3,kir:5}. This universal ``order of
nature" and evolution law unifies the extended versions of all
(correct) conservation laws, symmetries, and postulated principles
(which are causally derived and realistically interpreted now).
Contrary to any unitary symmetry, the universal symmetry of
complexity is \emph{irregular} in its structure, but always
\emph{exact} (never ``broken"). Its ``horizontal" manifestation
(at a given level of complexity) implies the actual, dynamic
symmetry between realisations, which are really taken by the
system, constituting the system dynamics (and evolution) and
replacing the abstract ``symmetry operators". Therefore the
conservation, or symmetry, of system complexity totally determines
its dynamics and explains the deep ``equivalence" between the
emerging, often quite dissimilar and chaotically changing system
configurations \cite{kir:3}.

Another, ``vertical" manifestation of the universal symmetry of
complexity is somewhat more involved and determines emergence and
development of different levels of complexity within a real
interaction process. System ``potentialities", or (real) power to
create new structure at the very beginning of interaction process
(before any actual structure emergence) can be universally
characterised by a form of complexity called \emph{dynamic
information} and generalising the usual ``potential energy"
\cite{kir:1,kir:3,kir:5}. During the interaction process
development, or structure creation, this potential, latent form of
complexity is progressively transformed into its explicit,
``unfolded" form called \emph{dynamic entropy} (it generalises
kinetic, or heat, energy). The universal \emph{conservation of
complexity} means that this important transformation, determining
every system dynamics and evolution, happens so that the sum of
dynamic information and dynamic entropy, or \emph{total
complexity}, remains unchanged (for a given system or process).
This is the absolutely universal formulation of the symmetry of
complexity, that includes the above ``horizontal" manifestation
and, for example, extended and unified versions of the first and
second laws of thermodynamics (i.e. conservation of energy and its
permanent degradation). It also helps to eliminate the persisting
(and inevitable) series of confusions around the notions of
information, entropy, complexity, and their relation to real
system dynamics in the unitary theory (thus, really expressed and
processed ``information" corresponds rather to a particular case
of our generalised dynamic entropy, see \cite{kir:1,kir:5} for
further details).

It is not difficult to show \cite{kir:1,kir:3,kir:5,kir:8} that
the natural, universal measure of dynamic information is provided
by the (generalised) action $\cal A$ known from classical
mechanics, but now acquiring a much wider, essentially nonlinear
and causally complete meaning applicable at any level of
complexity. One obtains then the universal differential expression
of complexity conservation law in the form of generalised
Hamilton-Jacobi equation for action ${\cal A} = {\cal A} (x,t)$:
\begin{equation}\label{eq:23}
\frac{{\Delta \cal A}}{{\Delta t}}\left| {_{x = {\rm const}} }
\right. + H\left( {x,\frac{{\Delta \cal A}}{{\Delta x}}\left| {_{t
= {\rm const}} ,t} \right.} \right) = 0\ ,
\end{equation}
where the \emph{Hamiltonian}, $H = H(x,p,t)$, considered as a
function of emerging space coordinate $x$, momentum $p = \left(
{{{\Delta \cal A} \mathord{\left/ {\vphantom {{\Delta A} {\Delta
x}}} \right. \kern-\nulldelimiterspace} {\Delta x}}} \right)\left|
{_{t = {\rm const}} } \right.$, and time $t$, expresses the
unfolded, entropy-like form of differential complexity, $H =
\left( {{{\Delta S} \mathord{\left/ {\vphantom {{\Delta S} {\Delta
t}}} \right. \kern-\nulldelimiterspace} {\Delta t}}} \right)\left|
{_{x = {\rm const}} } \right.$ (note that the discrete, rather
than usual continuous, versions of derivatives and variable
increments here reflect the naturally quantized character of
unreduced complex dynamics \cite{kir:1,kir:3,kir:5,kir:8}). Taking
into account the dual character of multivalued dynamics, where
every structural element contains permanent transformation from
the localised, ``regular" realisation to the extended
configuration of the intermediate realisation of generalised
wavefunction and back (Sect. 2), we obtain the universal
Schr\"odinger equation for the wavefunction (or distribution
function) ${\mit \Psi} (x,t)$ by applying the causal, dynamically
derived quantization procedure
\cite{kir:1,kir:3,kir:5,kir:8,kir:9} to the generalised
Hamilton-Jacobi equation (\ref{eq:23}):
\begin{equation}\label{eq:24}
\frac{{\partial {\mit \Psi} }}{{\partial t}} = \hat H\left(
{x,\frac{\partial }{{\partial x}},t} \right){\mit \Psi}\ ,
\end{equation}
where ${\cal A}_0$ is a characteristic action value (equal to
Planck's constant at quantum levels of complexity) and the
Hamiltonian operator, $\hat H$, is obtained from the Hamiltonian
function $H = H(x,p,t)$ of equation (\ref{eq:23}) with the help of
causal quantization (we also put here continuous derivatives for
simplicity).

Equations (\ref{eq:23})-(\ref{eq:24}) represent the universal
differential expression of the symmetry of complexity showing how
it directly determines dynamics and evolution of any system or
interaction process (they justify also our use of the Hamiltonian
form for the starting existence equation, Sect. 2). This
universally applicable Hamilton-Schr\"odinger formalism can be
useful for rigorous description of any complex network and its
separate devices, provided we find the \emph{truly complete}
(dynamically multivalued) general solution to particular versions
of equations (\ref{eq:23})-(\ref{eq:24}) with the help of
unreduced EP method (Sect. 2).

We have demonstrated in that way the fundamental, analytical basis
of description and understanding of complex (multivalued) dynamics
of real communication networks and related systems, which can be
further developed in particular applications in combination with
other approaches. The main \emph{practical proposition} of the
emerging intelligent communication paradigm is to open the way for
the \emph{free, self-developing structure creation} in
communication networks and tools with strong interaction
(including self-developing internet structure, intelligent search
engines, and distributed knowledge bases). The liberated,
autonomous system dynamics and structure creation, ``loosely"
governed by the hierarchy of system interactions as described in
this report, should essentially exceed the possibilities of usual,
deterministic programming and control.

%
%


\begin{thebibliography}{10}
\bibitem {kir:8}
Kirilyuk, A.P.: Quantum field mechanics: Complex-dynamical
completion of fundamental physics and its experimental
implications. Nova Science, New York (accepted for publication).
E-print physics/0401164
\end{thebibliography}
\end{document}